\documentclass[aps,prb,twocolumn,showpacs,superscriptaddress,floatfix]{revtex4-1}

\usepackage[titletoc,toc,title]{appendix}
\usepackage{amsmath,amssymb}
\usepackage[version=3]{mhchem} 
\usepackage{graphicx}
\usepackage{subfigure}
\usepackage{color}
\usepackage{rotating}
\usepackage{bm}
\usepackage{setspace}
\usepackage{hyperref}
\usepackage{float}
\usepackage{soul}
\usepackage{color}
\usepackage{braket}
\usepackage{xfrac}
\usepackage{changes}
\usepackage{ulem}

\hypersetup{
	colorlinks=true,
	urlcolor= blue,
	citecolor=blue,
	linkcolor= blue,
}

\renewcommand{\vec}[1]{\bm{#1}}

\newcommand{\EF}{E_\mathrm{F}} 
\newcommand{\etal}{\textit{et al.}}

\newcommand{\angstrom}{\textup{\AA}}

\begin{document}
	
	\preprint{APS/123-QED}
	
	\title{
		Electronic and Optical properties of transition metal dichalcogenides under symmetric and asymmetric field-effect doping		
	}
	\author{Peiliang Zhao}
	\affiliation{Key Laboratory of Artificial Micro- and Nano-structures of Ministry of Education and School of Physics and Technology,
		Wuhan University, Wuhan 430072, China}
	\affiliation{Institute for Molecules and Materials, Radboud University, Heyendaalseweg 135, 6525AJ Nijmegen, The Netherlands}
	\author{Jin Yu}
	\affiliation{Institute for Molecules and Materials, Radboud University, Heyendaalseweg 135, 6525AJ Nijmegen, The Netherlands}
	\author{H. Zhong}
	\affiliation{Key Laboratory of Artificial Micro- and Nano-structures of Ministry of Education and School of Physics and Technology,
		Wuhan University, Wuhan 430072, China}
	\author{M. R\"osner}
	\affiliation{Institute for Molecules and Materials, Radboud University, Heyendaalseweg 135, 6525AJ Nijmegen, The Netherlands}
	\author{Mikhail I. Katsnelson}
	\affiliation{Institute for Molecules and Materials, Radboud University, Heyendaalseweg 135, 6525AJ Nijmegen, The Netherlands}
	\author{Shengjun Yuan}
	\email{s.yuan@whu.edu.cn}
	\affiliation{Key Laboratory of Artificial Micro- and Nano-structures of Ministry of Education and School of Physics and Technology,
		Wuhan University, Wuhan 430072, China}

	\begin{abstract}
		
		Doping via electrostatic gating is a powerful and widely used technique to tune the electron densities in layered materials.
		The microscopic details of how these doping strategies affect the layered material are, however, subtle and call for careful theoretical treatments. 
		The external gates do not just increase the Fermi level in the system, but also generate external electric fields which affect the layered material as well.
		As a result, the electron densities within the system can redistribute and might thereby affect the electronic band structure in a non-trivial way.
		Theoretical descriptions via rigid shifts of the Fermi level can, therefore, be highly inaccurate.
		Using semiconducting monolayers of transition metal dichalcogenides (TMDs) as prototypical systems affected by electrostatic gating, we show that the electronic and optical properties change indeed dramatically when the gating geometry is properly taken into account. 
		This effect is implemented by a self-consistent calculation of the Coulomb interaction between the charges in different sub-layers within the tight-binding approximation. Thereby we consider both single- and double-sided gating.
		Our results show that, at low doping levels of $10^{13}$ cm$^{-2}$, the electronic bands of monolayer TMDs shift rigidly for both types of gating, and subsequently undergo a Lifshitz transition. 
		When approaching doping levels of $10^{14}$ cm$^{-2}$, the band structure changes dramatically, especially in the case of single-sided gating where we find that monolayer \ce{MoS_2} and \ce{WS_2} become indirect gap semiconductors. 
		The optical conductivities calculated within linear response theory also show clear signatures of these doping-induced band structure renormalizations.
		Our numerical results based on light-weighted tight-binding models indicate the importance of charge screening in doped layered structures, and pave the way for further understanding gated super-lattice structures formed by multilayers with extended Moir\'{e} patterns.
		
	\end{abstract}
	
	\maketitle  
	\section{\label{sec:level1}INTRODUCTION}
	
	Semiconducting transition metal dichalcogenides (TMDs) monolayers (\ce{MX2} with M=Mo, W and X=S, Se) ~\cite{novoselov_two-dimensional_2005} are direct gap semiconductors with optical gaps in the visible and near-infrared spectral range \cite{splendiani_emerging_2010, mak_atomically_2010, wang_electronics_2012}.
	Due to a variety of electronic\cite{fiori_electronics_2014, wang_electronics_2012} , optical\cite{xia_two-dimensional_2014, mak_photonics_2016, liu_optical_2014} and valleytronic\cite{mak_control_2012, zeng_valley_2012, xiao_coupled_2012} properties, 
	TMDs are expected to be utilized in various electronic and optoelectronic devices ~\cite{c.ferrari_science_2015,wang_electronics_2012,mak_photonics_2016} such as field effect transistors ~\cite{podzorov_high-mobility_2004, radisavljevic_single-layer_2011, fang_high-performance_2012}, photodetectors ~\cite{yin_single-layer_2012, lee_mos2_2012, gourmelon_ms2_1997, koppens_photodetectors_2014}, modulators ~\cite{sun_optical_2016, li_single-nanoparticle_2017} and electroluminescent devices \cite{carladous_light_2002, sundaram_electroluminescence_2013}. 
	When stacked with other two-dimensional (2D) materials such as graphene or hexagonal boron nitride, the resulting heterostructures can show highly sensitive photodetection and gate-tunable persistent photoconductivity at room temperature ~\cite{roy_graphenemos2_2013,georgiou_vertical_2013, bertolazzi_nonvolatile_2013,bernardi_extraordinary_2013}. 
	Upon electron-doping using ionic liquid gates a plethora of phases ranging from semimetallic, to metallic and superconducting regimes can be probed in TMDs and charge and magnetic order can be induced~\cite{ye_superconducting_2012, zhang_magnetic_2007,tsen_structure_2015,ritschel_orbital_2015,piatti_multi-valley_2018}.
	This field-effect induced doping can accumulate up to $10^{14}$ electrons per cm$^{2}$ in the layer~\cite{prete_iii-v_2019, lieb_ionic_2019,zheliuk_josephson_2019}, which can correspondingly affect all of these correlation effects.
	
	Here, we explore the electronic and optical properties of TMDs under the influence of those external electric fields resulting from asymmetric one- and recently realized  symmetric two-sided~\cite{zheliuk_josephson_2019} gates. 
	Based on a multi-orbital tight-bind model \cite{cappelluti_tight-binding_2013, Jose_lectronic_2016}, we implement a method ~\cite{mccann_asymmetry_2006, castro_biased_2007, avetisyan_electric_2009} to self-consistently calculate the induced charge (re)distribution within the different sub-layers of TMD monolayers, which is here especially accurate to describe the low-energy valleys of the valence and conduction bands.
	
	Our results show that for low doping levels of up to about $10^{13}$ cm$^{-2}$ the electronic band structure is just slightly renormalized independently of the gating geometry. Below the Lifshitz transition the electronic and optical features are very similar to the one obtained from simple rigid-shifts of the Fermi level. Upon the Lifshitz transition we, however, find clear optical features of the gate-induced band structure renormalizations.
	Upon further increasing the doping level to about $10^{14}$ cm$^{-2}$ the lowest conduction and upmost valence bands change remarkably under one-sided gating. 
	In the cases of MoS$_2$ and WS$_2$ these changes can yield direct-to-indirect band gap transitions.
	In contrast to the asymmetric gating, the symmetric gating geometry does not induce a direct-to-indirect band gap transitiion like that, but still strongly renormalizes the electronic dispersion.
	As the band structure renormalizations may lead to changes of the materials’ optical properties, we then present the calculated optical conductivities of TMDs based on the linear-response theory, to characterize the doping-induced effects. 
	Finally, we discuss the differences between local field-induced screening effects investigated here and non-local ones arising from internal polarizations as described by high level GW theories.
	
	The paper is organized as follows: In Sec.\ref{sec:THEORETICAL} the TMD tight-binding model is introduced together with the self-consistent calculation scheme to capture the externally induced band structure renormalization resulting from the one- or two-sided gates. In Sec.\ref{sec:Structure} and Sec.\ref{sec:optical} the electronic structure and optical spectra of gated TMDs are studied for low and high electron-doping regimes followed by a brief discussion and conclusion in Sec.\ref{sec:conclusion}. 
	
	\begin{figure}[t]
		\centering
		\subfigure[\quad One-sided Gate]
		{
			\includegraphics[width=0.43\columnwidth] {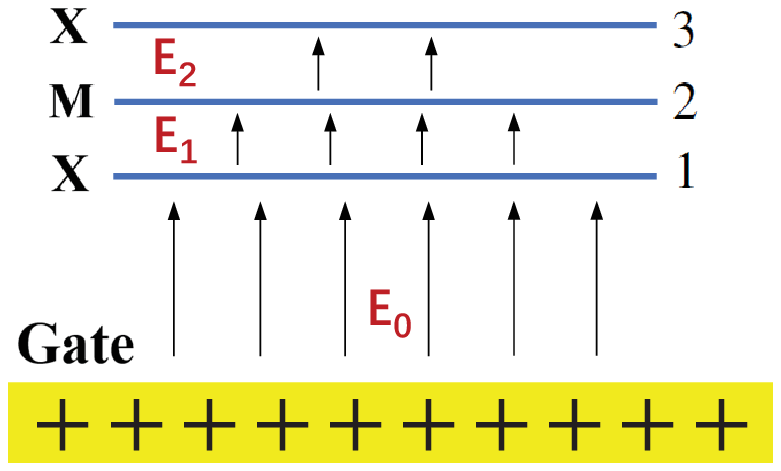}
		} 
		\subfigure[\quad Two-sided Gates]
		{
			\includegraphics[width=0.43\columnwidth] {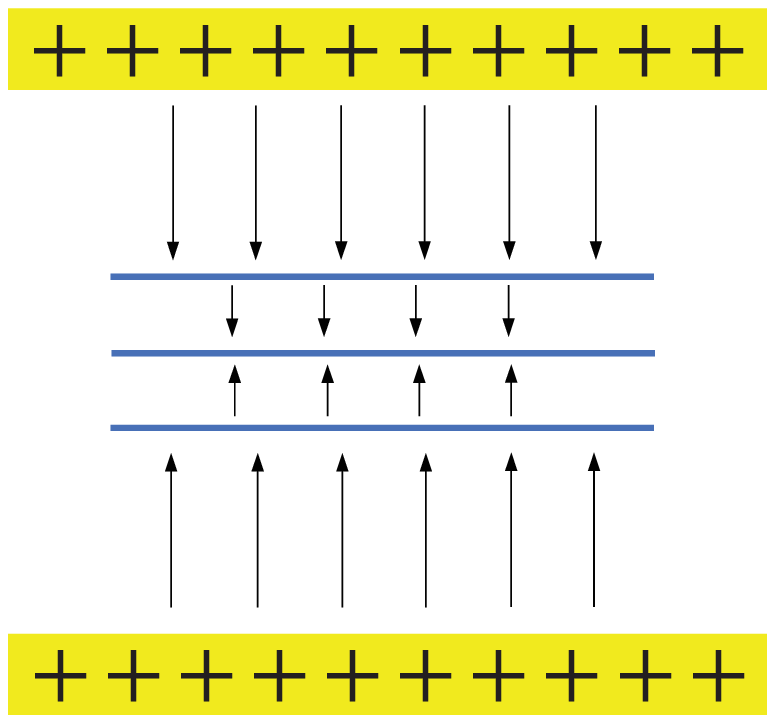}
		}\quad
		\caption{%
			Sketches of a TMD monolayer with positively charged gates (a): One-sided gate;  (b): Two-sided gates. 
			A total excess density of $n=n_1+n_2+n_3$ is induced, with $n_1$ ($n_3$) and $n_2$ being the excess densities on the bottom (top) \ce{X} and middle M sublayer, respectively.
		}	
		\label{fig:model&BD}
	\end{figure}
	
	\section{model and methods}
	\label{sec:THEORETICAL}
	
	The TMD crystals in our simulations are monolayers in the 2H-phase, which are formed by a top (X$^{\rm T}$) and a bottom (X$^{\rm B}$) chalcogen sub-layer, and a transition metal middle-plane \ce{M}. We model the undoped electronic bandstructure utilizing a long-range tight-binding model consisting of five \ce{M} \textit{d} orbitals and three \ce{X} \textit{p} orbitals \cite{cappelluti_tight-binding_2013, Jose_lectronic_2016}. 
	The Hilbert space is defined by
	\begin{equation}
	\label{eqn:basis_shi}
	\hat{\psi}^{\dagger}_{R_i}=
	\left[\hat{p}^{\rm T \dagger}_{R_i,\rm \alpha}, \,
	\hat{d}^{\dagger}_{R_i,\rm \beta}, \,
	\hat{p}^{\rm B \dagger}_{R_i,\rm \alpha}\right],
	\end{equation}
	where $\textit{d}_{R_i,\rm \beta}$ creates an electron in one of the d orbitals $\beta \in \{z^2, xy, x^2-y^2, xz, yz\}$ of the M atom and $\textit{p}_{R_i,\rm \alpha}$ creates an electron in one of the p orbitals  $\beta \in \{x, y, z\}$ of the X atoms in the $R_i$-unit cell.
	Using this basis the tight-binding Hamiltonian is given by:
	\begin{eqnarray}
	H_0
	&=&
	\sum_{k}
	\phi_{k}^\dagger
	\hat{H}_{0,k}
	\phi_{k},
	\label{ham}
	\end{eqnarray}
	where $\phi_{k}$ is the Fourier transform of $\psi_{R_i}$ in momentum space.
	The Hamiltonian  $\hat{H}_{0, k}$  can be written as (we omit the index ${k}$ for simplicity from now on):
	\begin{eqnarray}
	\hat{H}_0
	&=&
	\left(
	\begin{array}{ccc}
	\hat{H}_{p^{\rm T},p^{\rm T}} & \hat{H}_{d,p^{\rm T}} & \hat{H}_{p^{\rm T},p^{\rm B}} \\
	\hat{H}_{d,p^{\rm T}}^\dagger & \hat{H}_{d,d} & \hat{H}_{d,p^{\rm B}} \\
	\hat{H}_{p^{\rm T},p^{\rm B}}^\dagger & \hat{H}_{d,p^{\rm B}}^\dagger &\hat{H}_{p^{\rm B},p^{\rm B}} 
	\end{array}
	\right).
	\label{hmatr11}
	\end{eqnarray}
	All involved lattice parameters are given in Table~\ref{tab:lattice}, where $a$ and $c$ are the in- and out-of-plane lattice constants, and $z$ is the sub-layer distance between the \ce{M} and \ce{X} planes. 
	The tight-binding parametrization is taken from Ref. \onlinecite{Jose_lectronic_2016}, which accurately reproduces the electronic dispersion and orbital characters of the lowest conduction band and upmost valence band. We neglect the effect of spin-orbit coupling since it will not drastically effect the charge-redistributions between the sub-layers and can be easily added afterwards via simple Russel-Saunders like approaches~\cite{liu_three-band_2013}.
	We also neglect possible geometric relaxations upon electron doping. Full ab initio calculations have shown that these are rather small\cite{brumme_first-principles_2015}.
	
	\begin{table}[t]
		\caption{\label{tab:lattice} Lattice parameters for the TMDs considered here. $a$ represents the in-plane lattice constant, $z$ is the distance between the \ce{M} and \ce{X} planes, and $c$ accounts for the distance between the \ce{M} layers.   }
		\begin{ruledtabular}
			\begin{tabular}{lccc}
				&    $a$        &    $c$     &    $z$ \\
				\hline
				MoS$_2$ & $3.160\:\angstrom$  & $12.28\:\angstrom$ & $1.586\:\angstrom$ \\
				WS$_2$  & $3.153\:\angstrom$  & $12.32\:\angstrom$ & $1.571\:\angstrom$ \\
				WSe$_2$ & $3.260\:\angstrom$  & $12.84\:\angstrom$ & $1.657\:\angstrom$ 
			\end{tabular}
		\end{ruledtabular}
	\end{table}
	
	Upon gating the TMD monolayer and applying an external electric field  additional (excess) electrons will accumulate within the monolayer\cite{ye_superconducting_2012, zhang_magnetic_2007,tsen_structure_2015,ritschel_orbital_2015}, as shown in Fig.~\ref{fig:model&BD}. 
	In detail, an asymmetric one-sided gate creates an uniform electric field $\mathrm{E}= ne/2 \epsilon_0\kappa$, where $n = n_1+n_2+n_3$ is the excess electron density, with $n_1$ ($n_3$) describing the excess electron density in the bottom (top) X-sublayer and $n_2$ the excess electron density on the middle M-sublayer. 
	The induced excess electrons $n_i$ redistribute as a reaction to this external gate field and create in turn uniform electric fields $\mathrm{E}_{i} \, (i=1,2)$ between the sub-layers with $\mathrm{E}_{1}=(n_2+n_3)e /\kappa \varepsilon_{0}$ and $\mathrm{E}_{2}=n_3 e /\kappa \varepsilon_{0}$.
	This effectively \emph{screens} the external gate field.
	
	In order to find the resulting distribution of these excess electron densities $n_i$, we make use of the self-consistent approach from Refs.~\onlinecite{mccann_asymmetry_2006, castro_biased_2007, avetisyan_electric_2009}, which has recently also been applied to multilayer phosphorene~\cite{li_electric-field_2018,li_tuning_2018,li_strain_2019}.
	Accordingly, $n_i$ create electrostatic potentials $\Delta_i$ ($i = 1, 2, 3$) in each sub-layer, which are given in the one-sided gating setup by
	\begin{align}
	\Delta_1(\vec{n}) &=  +\gamma (n_2+n_3),\label{eq:4}\\
	\Delta_3(\vec{n}) &=  -\gamma n_3, \label{eq:5}
	\end{align}
	where $\gamma= e^2z/\epsilon_0 \kappa$, $\vec{n} = (n_1, n_2, n_3)$, $\epsilon_0$ is the vacuum permittivity and $\kappa$ is the dielectric constant. Here, we choose \ce{SiO2} as the gates and set $\kappa=(1.0+\kappa [{ \ce{SiO2}}])/2$. 
	In both one- or two- sided geometries, the electrostatic potential $\Delta_2$ in the \ce{M} middle layer is set to  zero.
	The one-sided gating geometry thus introduces a sub-layer asymmetry. 
	For the two-sided gating setup, positive charge carriers are introduced equally in the outmost X sub-layers, retaining the mirror symmetry with respect to the M-plane. 
	As a result we find
	\begin{align}
	\Delta_1(\vec{n}) = \Delta_3(\vec{n}) = \frac{\gamma}{2} n_2.
	\end{align}
	The full Hamiltonian in the presence of the external electric field is thus given by
	\begin{eqnarray}
	\hat{H}(\vec{n})
	&=&
	\hat{H}_0 + 
	\begin{pmatrix}
	\Delta_1(\vec{n}) && \\
	& \ddots &\\
	&& \Delta_3(\vec{n})
	\end{pmatrix},
	\label{hmatr12}
	\end{eqnarray}
	where $\Delta_1(\vec{n})$ and $\Delta_3(\vec{n})$ are  $3 \times 3$ diagonal matrices in the basis of the chalcogen $p$ orbitals with diagonal elements defined in Eqs. (\ref{eq:4},\,\ref{eq:5}).
	Using the density of states (DOS) 
	\begin{equation}
	\rho (\varepsilon) =\frac{1}{2\pi} \sum_{n=0}^{10} \int_{\mathbb{BZ}} \delta [\varepsilon-E_n(k)] dk,
	\end{equation}	
	and the eigenfunctions of $\hat{H}(\vec{n} )$  
	\begin{equation}\label{eqn:eigenfunction}
	\phi_{k}^{\dagger} =
	\left[p^{\rm T \dagger}_{k,\rm \alpha}, \,
	d^{\dagger}_{k,\rm \beta}, \,
	p^{\rm B \dagger}_{k,\rm \alpha}\right],
	\end{equation}
	the sub-layer DOS are naturally given by
	\begin{align*}
	\rho_1 (\varepsilon) &= \frac{1}{2\pi} \sum_{m=0}^{10} \int dk \, \delta [\varepsilon-E_m(k)]  \sum_\alpha \left|p^{\rm T \dagger}_{k,\rm \alpha}\right|^2, \\
	\rho_2 (\varepsilon) &= \frac{1}{2\pi} \sum_{m=0}^{10} \int dk\, \delta [\varepsilon-E_m(k)]  \sum_\beta \left|d^{ \dagger}_{k,\rm \beta}\right|^2 , \\
	\rho_3 (\varepsilon) &= \frac{1}{2\pi} \sum_{m=0}^{10} \int dk\, \delta [\varepsilon-E_m(k)]   \sum_\alpha \left|p^{\rm B \dagger}_{k,\rm \alpha}\right|^2.
	\end{align*}	
	We vary $\vec{n}$ until we find a self-consistent solution with $n_i = \int_{0}^{\Delta \EF}\rho_i \left( \varepsilon \right) d\varepsilon $ using $\Delta \EF = \EF(n) - \EF(0) $ and $\EF(n)$ being the doping-dependent Fermi level.
	
	\section{Electronic Band Structure}
	\label{sec:Structure}
	
	The low-energy band structure of TMD monolayers around the band gap is mostly characterized by two valleys at the $K$ and around the $Q$ points in the conduction band and by two valleys at the $\Gamma$ and $K$ points in the valence band (see, e.g., Ref.~\onlinecite{Jose_lectronic_2016}). In the conduction band the $K$ valley is predominately of $d_{z^2}$ character, while the valley around the $Q$ point results from hybridized X $p$ orbitals and M $d$ orbitals\cite{Jose_lectronic_2016}.
	Upon electron doping the $K$ and $Q$ pockets become successively occupied. 
	Due to the doping-induced potentials $\Delta_i$ they also shift in energy, which renormalizes the electronic dispersion in contrast to rigid-shifts of the Fermi level.
	
\begin{figure}[t]
		\centering
		\includegraphics[width=1.0\columnwidth] {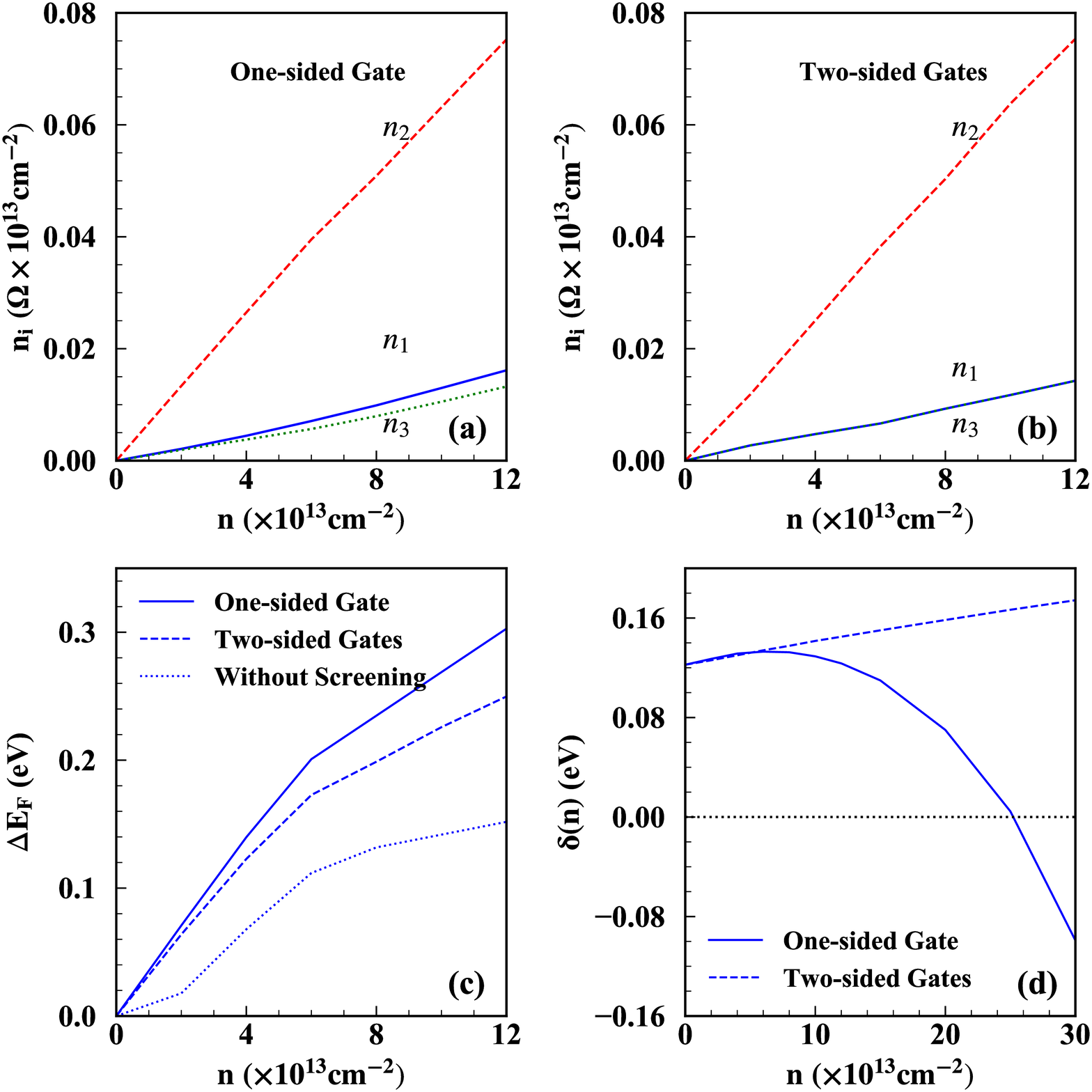}
		\caption{%
			Partial densities $n_i$ in (a) one-sided and (b) two-sided gate geometries in dependence on the total excess density $n$ for electron-doped monolayer MoS$_2$, where $\Omega$ is the area of a unit cell.
			(c) $\Delta \EF$ as a function of $n$ in the presence of both gating geometries and for simple rigid shifts of the Fermi level. The solid and dashed lines correspond to one-sided and two-sided gating geometries, respectively, and the dotted line to rigid-shifts of the Fermi level. 
			(d) K/Q valley detuning $\delta(n)$ as a function of $n$ for both gating geometries.
		}
		\label{fig:fermi_level}
	\end{figure}

	Since these potentials $\Delta_i$ are defined by the self-consistently calculated partial excess electron densities $n_i$, we start by analyzing the latter as a function of the total doping level $n$, as shown for MoS$_2$ in  Fig.~\ref{fig:fermi_level} (a) and (b) for both gating setups.
	The excess electrons are distributed unevenly between the sub-layers with the main contribution on the central Mo-layer and smaller contributions on the chalcogen atoms. 
	Upon increasing the doping level $n$, electrons get further localized on the \ce{Mo} layer. 
	For the one-sided gating, the excess electron density on the first layer, which is closest to the gate, is slightly larger than those on the third layer.
	Due to the symmetric $\Delta_1 = \Delta_3$ in the two-sided gating geometry, $n_1$ and $n_3$ are also symmetrically distributed to the chalcogen sub-layers. 
	For $n = 1.0 \times 10^{14}\,$ cm$^{-2}$ we find that about $73\%$ of doping electrons are located on the \ce{Mo} layer, while $15\%$ and $12\%$ are accumulated at the bottom and top \ce{S} layers in the one-sided gate geometry.
	In the case of the two-sided gates, about $73\%$ of the excess electrons are localized at the \ce{Mo} layer, and the two \ce{S} layers each hold $13.5\%$, respectively.
	This enhanced inhomogeneity in the layer-resolved charge distribution in TMD monoalyers was also reported in %
	a similar ab initio study for single-side doping geometry by Brumme \etal ~\cite{brumme_first-principles_2015}.

	\begin{figure*}[t]
		\centering
		\includegraphics[height= 0.5\textheight]{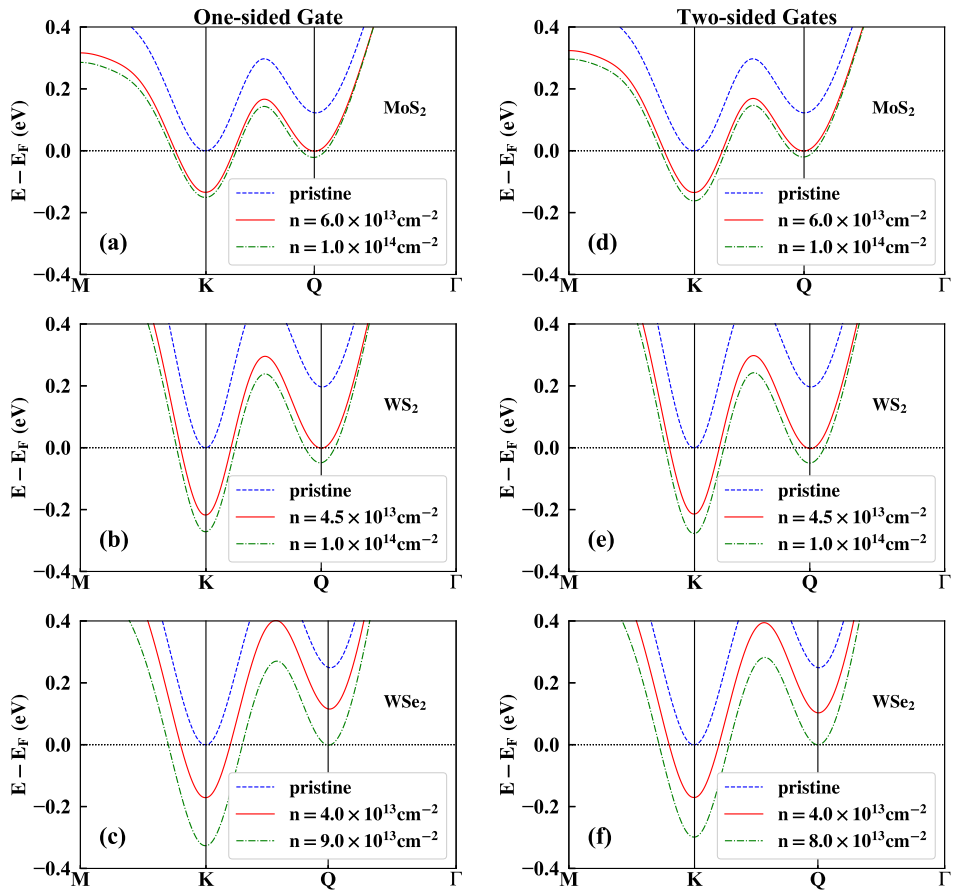}	  	
		\caption{Band structures of electron-doped TMDs in low doping regime  $n \leq 10^{14} \,$ cm$^{-2}$. 
			Figures (a) (b) (c) are for one-sided gate geometries and (d) (e) (f) show the two-sided gate geometries. 
			The solid vertical lines around $Q$ mark the $Q$ valley positions in green dash-dotted lines.
		}
		\label{fig:low_doping}
	\end{figure*}

	Fig.~\ref{fig:fermi_level} (c) shows the change in the Fermi level $\Delta E_F(n)$ as a function of gate-induced doping for MoS$_2$ for both gating geometries and for simple rigid-shifts (i.e. without any screening). 
	For all of these scenarios $E_F(n)$ naturally increases with electron doping $n$. 
	The two gate geometries behave rather similar, with the only difference of a slightly reduced shift in the case of the two-sided gates. 
	In contrast, the rigid-shift approximation strongly underestimates the shifts of the Fermi level due to missing renormalizations of the band structure.
	From the comparison to the rigid-shift scenario, we understand that the doping-induced band structure renormalizations are the strongest in the one-sided gating geometry.
	This effect is slightly reduced in the symmetric two-sided gating geometry, but still non-negligible.
	Next to these renormalization-induced effects, we clearly see a reduced enhancement of $\Delta E_F(n)$ for $n > 6 \times 10^{13}\,$cm$^{-2}$.
	This is attributed to the occupation of the $Q$ valleys, which induces a Lifshitz transition and slows down the Fermi-level shift, which we discuss in detail in the following. 
	

	\begin{figure*}[t]
		\centering
		\includegraphics[height= 0.5\textheight]{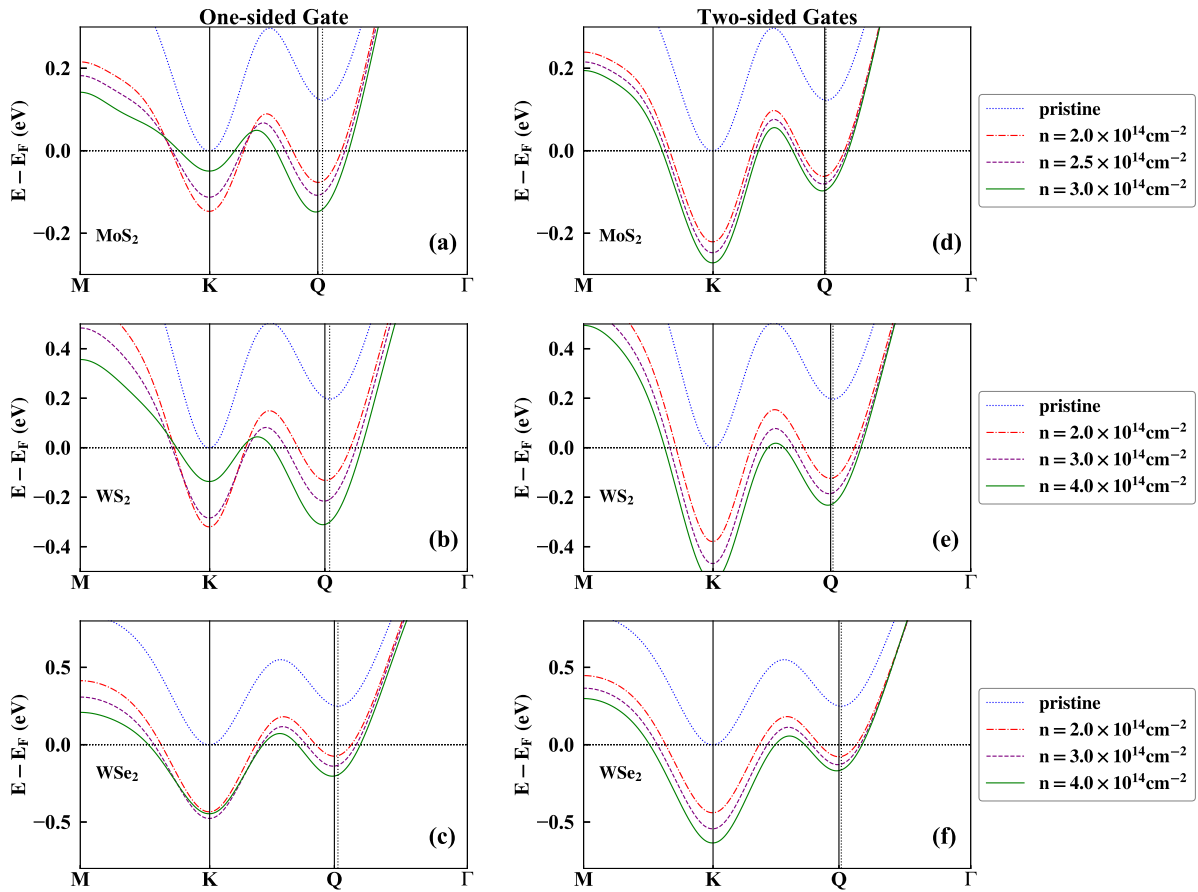}	  	
		\caption{Band structures of electron-doped TMDs in high doping regime  $2.0 \times 10^{14}\,$ cm$^{-2}$  $\leq  n \leq 4.0 \times 10^{14} \,$ cm$^{-2}$. 
			Figures (a) (b) (c) are for one-sided gate geometries and (d) (e) (f) show the two-sided gate geometries. 
			The vertical dotted lines mark the $Q$ valley positions in BZ for pristine TMDs, while the solid vertical lines around $Q$ mark the $Q$ valley positions in green solid lines.
		}
		\label{fig:high_doping}
	\end{figure*}

	To this end we analyze the band structures of MoS$_2$, WS$_2$, and WSe$_2$ for different doping levels and for both gating geometries in Fig.~\ref{fig:low_doping}.
	Overall, all materials behave rather similar for the depicted ``low'' electron doping regimes ($n \leq 1.0 \times 10^{14} \,$ cm$^{-2}$), in which mostly the $K$ valley gets occupied and doping-induced band structure renormalizations are rather small.
	From this we also see that the changes in the Fermi levels (as measured from the bottom of the $K$ valley) are rather large as long as just $K$ is occupied. 
	As soon as $Q$ gets occupied as well the shift in the Fermi level slows down drastically.
	This results from the different orbital characters defining the $Q$ valley.
	While the $K$ valley is mostly of $d_{z^2}$ character, the $Q$ pocket results from a hyribdization of all (involved) orbitals. 
	Thus, the self-consistently calculated potentials $\Delta_{1/3}$ which mostly act on $p$ orbitals have a much stronger effect as soon as the $Q$ pocket gets occupied so that the renormalization effects are enhanced.
	
	As the $Q$ valleys get populated, TMDs undergo a Lifshitz transition that reconstructs the Fermi surface\cite{shkolnikov_valley_2002, rycerz_valley_2007, xiao_valley-contrasting_2007,liu_evidence_2010, schonhoff_interplay_2016}. Six new Fermi pockets centered around $Q/Q^{\prime}$ appear in the BZ and the Fermi surface topology changes drastically as shown in Fig.~\ref{fig:FMstructure}.
	While the gating geometry in this low doping regime does not affect MoS$_2$ and WS$_2$, it is important for WSe$_2$.
	The critical electron doping level that occupies the $Q$ valleys is clearly dependent on the gating setup, as shown in Fig.~\ref{fig:low_doping} (c) and (f). 
	With the one-sided gate, the Lifshitz transition happens at $n = 9.0\times10^{13}$ cm$^{-2}$, and for the two-sided gates, the corresponding electron doping level is $n = 8.0\times10^{13}$ cm$^{-2}$. 
	While these critical doping levels are the same in MoS$_2$ and WS$_2$ for both gating setups.

	
	In Fig.~\ref{fig:high_doping} we show the corresponding band structures for high doping densities. 
	Here, in the one-sided gate geometry strong band structure renormalizations are observed, which cannot be described by simple rigid-shifts of the Fermi level.
	For MoS$_2$ and WS$_2$ [Fig.~\ref{fig:high_doping} (a) and (b)] these renormalizations can shift the conduction band edge from $K$ to $Q$, resulting in a direct-to-indirect band-gap transition consistent with previous DFT and numerical results~\cite{brumme_first-principles_2015,erben_excitation-induced_2018}. 
	For \ce{WSe_2}, much higher electron doping densities are needed to realize such a transition.
	%
	%
	Another important characteristic in the high-doping regime is the renormalization of the $K$ valley as a function of electron density $n$. 
	In contrast to the low-doping regime, the pocket around the $K$ point shifts upward (with respect to the $Q$ valley) when the doping density increases.
	In the low-doping regime, the $K$ pocket shifts slowly down with increasing $n$. 
	%
	%
	Due to the symmetric $\Delta_1 = \Delta_3$ in the double-sided gate geometry, inter-valley renormalizations in the high-doping regime are strongly reduced in comparison to the single-sided gate.
	In this case, the shifts of the conduction band are however enhanced, as shown in Fig.~\ref{fig:high_doping} (e) and (f).
	In addition, the $Q$ valley is moving towards the $K$ valley within the BZ with increasing $n$, as also observed in the low-doping regime. Through high electron doping we can thus tune the Fermi surface topology and the relative alignment between the $K$ and $Q$ valleys.

	The quantitative differences between the considered TMDs result  from an interplay between the different chalcogen plane separations $z$ as listed in Tab \ref{tab:lattice}  and M $d$-$d$ and M $d$ - X $p$ orbital hybridization. These properties control the band gap, the $K/Q$ valley splitting, and eventually also the responses to the doping levels and the doping geometries. While a quantitative disentanglement of these effects is hard to achieve due to the applied self-consistent scheme, we can, however, qualitatively compare the different characteristics.
	Transition metal contribution: The chalcogen plane separation in WS$_2$ is just slightly smaller than in MoS$_2$ yielding a possibly negligible effect. The main difference between MoS$_2$ and WS$_2$ is thus the orbital hybridization, which we can be estimated from the orbital admixtures of the different valleys at the high symmetry points (see Tab. $3$ from Ref.~\onlinecite{Jose_lectronic_2016}). Especially in the valence and conduction $K$ valleys, we see that in WS$_2$ the relative transition metal $d$ orbital contribution is smaller and the S $p$ contribution is larger than in MoS$_2$. Hence, the slightly stronger doping effect in WS$_2$ compared to MoS$_2$ are likely due to the enhanced S $p$ admixtures in the relevant valleys.
	Chalcogen contribution: From a comparison of WS$_2$ and WSe$_2$, we find a significantly enhanced chalcogen layer separation in WSe$_2$, while the chalcogen $p$ orbital contributions in the relevant valleys in WSe$_2$ is smaller compared to WS$_2$. 
	The enhanced chalcogen layer separation in WSe$_2$ is thus responsible for the enhanced doping effects in WSe$_2$ compared to WS$_2$ at low doping levels, which mostly occupies the $K$ valley, so that $d$-$p$ orbital hybridization plays only a minor role.
	At high doping levels, however, WSe$_2$ is less affected compared to WS$_2$ since here the $Q$ valleys get (strongly) occupied, so that the increased $d$-$p$ hybridization in WS$_2$ becomes important and renders WS$_2$ more prone to electron doping.
	Our self-consistent gating-induced doping description based on a multi-orbital tight-binding model thus reproduces the trends found in numerically more-demanding ab initio calculations for the single-sided gate geometry\cite{brumme_first-principles_2015} and demonstrates how different the resulting renormalizations are in a symmetric gating setup.
	
	\section{Optical spectroscopy}
	\label{sec:optical}	
	
	Now, we turn to the effects of the gate-induced doping to the optical conductivity within linear response theory, which we calculate by using the Kubo formula \cite%
	{Ishihara1971} as implemented within our TBPM code\cite{YRK10} (omitting the Drude contribution
	at $\omega =0$)
	\begin{eqnarray}
	\sigma \left( \omega \right) &=&\lim_{\epsilon \rightarrow \infty }\frac{%
		e^{-\beta \omega }-1}{\omega \mathcal{A} }\int_{0}^{\infty }e^{-\epsilon t}\sin
	\omega t  \notag  \label{gabw2} \\
	&&\times 2~\text{Im}\left\langle \varphi |f\left( H\right) J\left(
	t\right) \left[ 1-f\left( H \right) \right] J|\varphi \right\rangle
	dt.  \notag \\
	&&  \label{Eq:OptCond}
	\end{eqnarray}%
	Here, $\beta =1/k_{B}T$ is the inverse temperature, $\mathcal{A}$ is the sample area, $f\left( H %
	\right) =1/\left[ e^{\beta \left( H-\mu _{F}\right) }+1\right] $
	is the Fermi-Dirac distribution operator and $\mu_{F}$ is the chemical potential. 
	In order to alleviate the effects of the finite time ($\tau$) in the numerical time integration, we adopt a Gaussian window of $10^{-\epsilon(t/T)^{2}}$ with $\epsilon = 2$ in Eq. (\ref{Eq:OptCond}).
	
\begin{figure}[t]
		\centering
		\includegraphics[width=0.48\columnwidth]{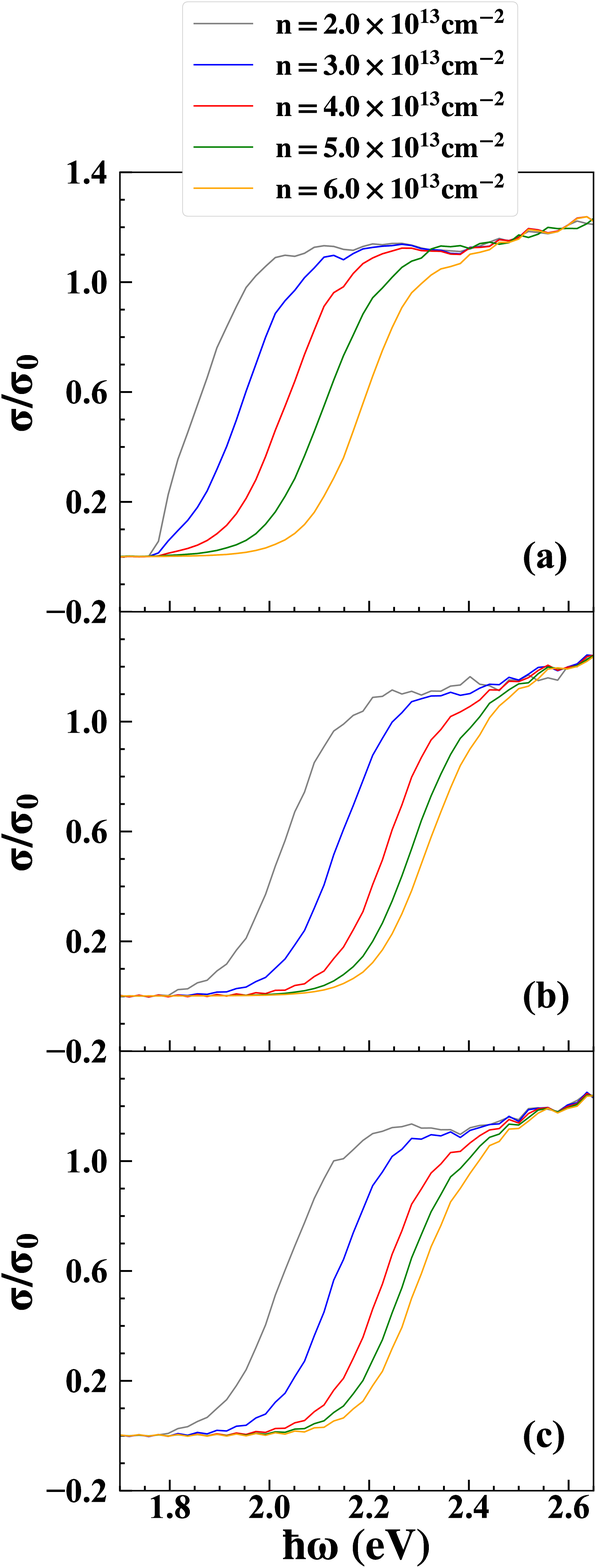}	  
		\includegraphics[width=0.48\columnwidth]{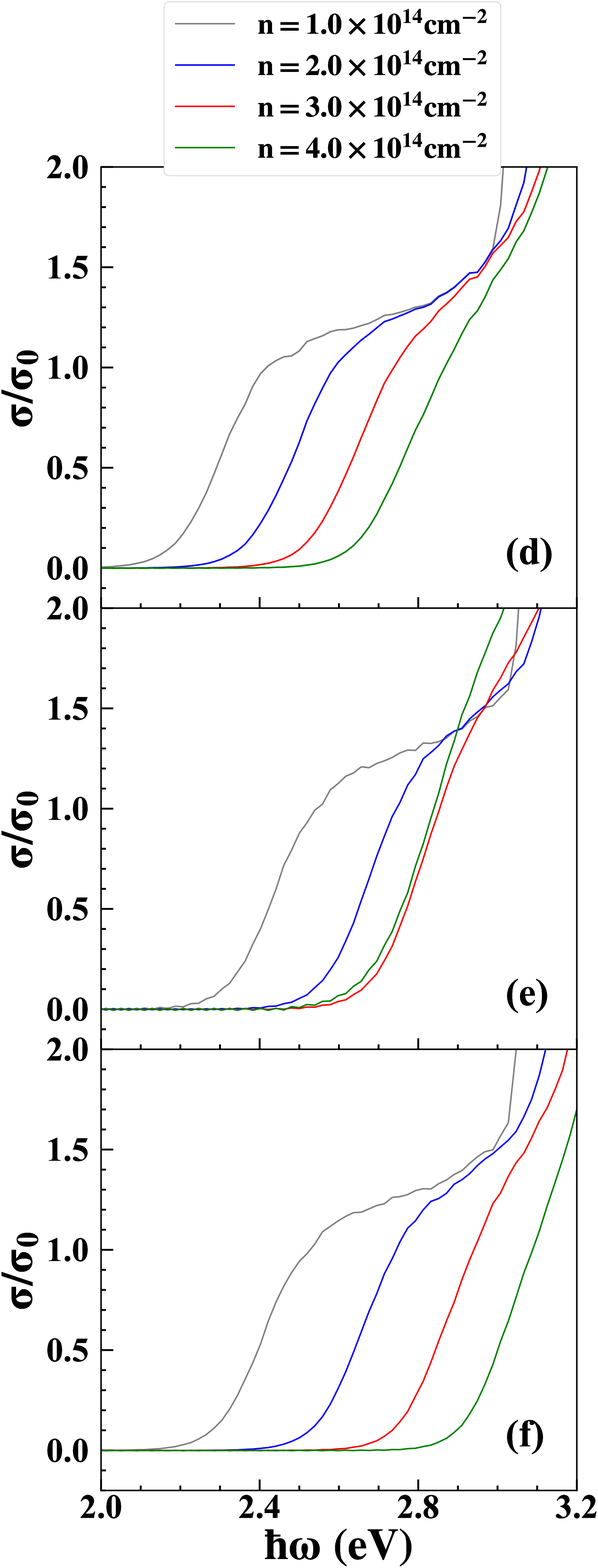}		
		\caption{Optical conductivities (in units of $\sigma_0 = \pi e^2/2h$) of WS$_2$ for different carrier densities $n$. (a)  (d) depict the optical conductivities for rigid shifts of the Fermi level. (b) (e) show the optical conductivities for  one-sided gate, and (c) (f) are for two-sided gate geometries, respectively.
		}
		\label{fig:ac}
	\end{figure}

	In Fig.~\ref{fig:ac} we show the resulting optical conductivities of WS$_2$ for (a,d) rigid-shifts of the Fermi level and for the (b,e) single and (c,f) dual gate geometries. 
	Fig.~\ref{fig:ac} (a) to (c) focus on the electron doping regime $2.0 \times 10^{13}\,$ cm$^{-2}$  $\leq  n \leq 6.0 \times 10^{13} \,$ cm$^{-2}$.
	The Lifshitz transition happens around $n=4.5 \times 10^{13}\,$ cm$^{-2}$ (from red to green).
	In general, the optical signals shown here (above the band gap) exhibit  blue shifts with increasing doping density corresponding to an increasing electronic band gap. 
	While this blue shift is unaltered by the Lifshitz transition, the one- and two-sided gating effects render the Lifshitz transition clearly visible in the form of reduced blue shifts.
	Fig.~\ref{fig:ac} (d) to (f) depict the optical conductivities in the high electron doping regime. 
	Here, the most obvious difference in these three doping scenarios is the "stop" of the blue shift in the asymmetric one-sided gate situation for  $n = 4.0 \times 10^{14} $ cm$^{-2}$.
	This is the doping level at which the direct-to-indirect band gap transitions occurs.
	The two-sided gate geometry scenario behaves similar to the rigid-shift situation, however, with strongly enhanced blue shifts. 
	These optical characteristics can thus be used to monitor the Lifshitz and the possible direct-to-indirect band-gap transitions. The latter should be seen only in the asymmetric gating situation.
	
	\section{Discussion And Conclusions}
	\label{sec:conclusion}
	
	
	We numerically studied the electronic and optical properties of electron-doped TMD monolayers by gating, considering well-known single- as well as novel double-sided gate geometries.
	The redistribution of the induced excess electron densities within the sub-layers of the TMD monolayers due to the applied gate field is self-consistently accounted for within a light-weighted tight-binding approach, resulting in considerably different excess electron densities distributions between the different geometries.
	Thereby static screening of the external gate field is intrinsically captured.
	The latter yield band structure renormalizations, prominently observed as relative shifts between the $K$ and $Q$ valleys in the conduction band.
	These renormalizations can have important consequences: On the one hand, they define the critical doping density corresponding to a Lifshitz transition, which drastically changes the Fermi surface topology by occupying the $Q$ pockets.
	On the other hand, these renormalization effects can be strong enough to induce a direct to indirect band-gap transition by shifting the $Q$ valley below the $K$ valley.
	Interestingly, the different TMDs exhibit opposite sensitivities at different doping levels. 
	While in the low-doping regime the WSe$_2$ band structure renormalizations are most sensitive to the doping (in comparison to WS$_2$ and MoS$_2$), the MoS$_2$ band structure is most sensitive to electron doping in the high-density regime. 
	These results are in-line with similar calculations based on full ab initio calculations applying density functional theory for the single-side gate geometry by Brumme \etal. \cite{brumme_first-principles_2015} and show that the commonly used rigid shift of the Fermi level in doped layered structures might miss important effects.
	Furthermore, we find that the double-sided gate geometry results in strongly different renormalizations of the electronic band structure at electron doping levels.
	Our tight-binding based approach can be straightforwardly generalized to structures with large supercells, such as twisted multilayers and their heterostructures, or Moir\'{e} patterns.
	These generally contain a large number of atoms which can easily exceed the computational limits of density functional calculations. 
	
	Regarding the consequences of the band structure renormalizations to TMDs, the possibility to perfectly align the $K$ and $Q$ valleys in \ce{MoS_2} and \ce{WS_2} can be useful to design valleytronic devices, as they might show an optimal performance when two or more valleys are available at similar energies but at different positions in momentum space~\cite{shkolnikov_valley_2002, rycerz_valley_2007, xiao_valley-contrasting_2007}.
	Also, the electron-phonon interaction in electron-doped TMDs depends strongly on which valleys of the conduction band are occupied, as the orbital characters of electronic states differ substantially in different valleys \cite{ge_phonon-mediated_2013}. Our results may help to explain the superconducting dome in gated TMDs \cite{liu_evidence_2010, schonhoff_interplay_2016, ye_superconducting_2012, lu_full_2018}  as well as details of charge-density ordering.
	
	Regarding optical properties, we clearly showed the existence of additional doping-induced features in the optical conductivity of TMD monolayers. 
	While these doping-induced features are similarly affected by the doping level in both, rigid-shift and gate-induced, scenarios, optical features at energies larger than the electronic band gap, certainly display changes induced by band-structure renormalizations, which are not present in the rigid-shift-like doping.
	Thus, optical probes can help to monitor both, the doping level and the correspondingly induced band renormalizations including changes to the Fermi surface topology, rendering them a powerful tool to characterize doping-induced effects.
	
	Here we, however, also see a clear shortcoming of our approach: The optical gaps increase upon doping, which result from increasing electronic band gaps in our calculations. 
	This contradicts the decreasing trends seen in $GW$-like calculations for increasing Fermi levels \cite{erben_excitation-induced_2018, steinhoff_2014}, which are experimentally verified upon optical doping \cite{PhysRevB.92.125417}.
	While these $GW$-like calculations take %
	explicitly exchange contributions including 
	the full long-range Coulomb interaction and the \emph{internal} screening of these interactions into account, our calculations describe local effects only.
	Thus, our calculations can be seen as a mean-filed treatment with local Coulomb interactions only, whereby the latter are successively reduced (screened) upon increasing doping concentration.
	And indeed, analogous LDA+U calculations also show an increasing electronic band gap upon \emph{decreasing} (screening) U.
	As discussed in Ref.~\onlinecite{roesner_hj_2016}, the band gap in semiconducting TMDs mostly results from hybridization effects between the transition metal $d$ orbitals.
	These hybridization effects are enhanced by long-range Coulomb interactions, which explains the increased band gap in $GW$ calculations for TMD monolayers in contrast to plain DFT calculations \cite{Haastrup_2018}.
	Thus, upon increasing screening due to increased Fermi levels the long-range Coulomb interaction is decreased, which decreases the $d$-orbital hybridization, which in turn must reduce the electronic band gap %
	and also renormalizes the overall band structure. For a detailed discussion of these doping-induced renormalizations we refer the reader to Ref.~\onlinecite{erben_excitation-induced_2018} , where the authors show the relative screening-induced modifications of the band structure due to Coulomb-hole and exchange contributions of the self-energy.
	Nevertheless, the local gate-field induced changes described here must be considered as well, which are so far missing in standard $GW$ calculations.
	Thus, in order to achieve a full quantitative description of gate-induced doping effects in layered materials, $GW$-like calculations are needed which take the external gate-field into account. 
	
	\begin{acknowledgments}
		\label{sec:acknowledgments}		
		
		We acknowledge helpful discussions with J. Silva-Guill{\'e} and G. Yu. 
		This work was supported by the National Key R$\&$D Program of China (Grant No.  2018FYA0305800). M.I.K. thanks financial support from JTC-FLAGERA Project GRANSPORT. Z.H thanks support from  the National Natural Science Foundation of China (Grant Nos. 11947218).
		Support by the Netherlands National Computing Facilities foundation (NCF), with funding from the
		Netherlands Organisation for Scientific Research (NWO), is gratefully acknowledged.
		Numerical  calculations  presented  in  this  paper  were  performed  partially on  the  Supercomputing Center of Wuhan University.
	\end{acknowledgments}
	
	\begin{appendix}	
		
		\section{Fermi surfaces reconstruction}
		
\begin{figure}[t]
			\centering
			\includegraphics[width=1.0\columnwidth] {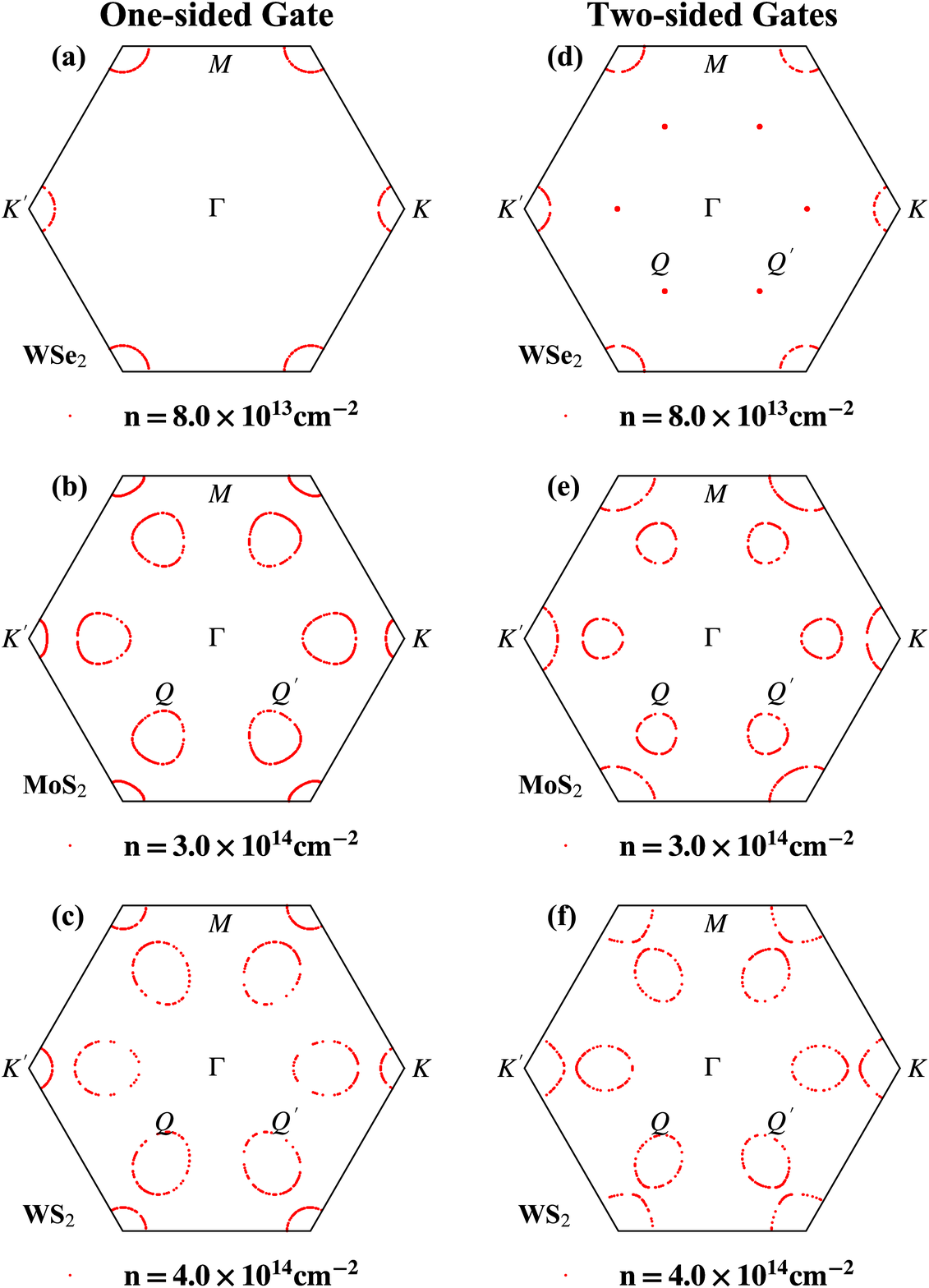}	  	
			\caption{The Fermi surfaces of TMD monolayers in dependence of electron doping density $n$.  	
			}
			\label{fig:FMstructure}
		\end{figure}

		The Fermi surfaces of TMD monolayers depend on the excess doping density $n$ and gating geometry. Fig~\ref{fig:FMstructure} depicts the Fermi surfaces of electron-doped TMDs. 
		Fig~\ref{fig:FMstructure} (a) to (c) correspond to the one-sided gate geometry and (d) to (f) to the two-sided gate setup. 
		As shown in Fig~\ref{fig:FMstructure} (a) and (d) at the same doping density $n = 8.0\times10^{13}$ cm$^{-2}$, six $Q/Q^{\prime}$ surfaces arise for the two-sided gate geometry, while in the one-sided gate case just the $K/K^{\prime}$ pockets are occupied, and no Lifshitz transition has occurred.
		If we account for band structure renormalizations within the single- and double-sided gate setups, one of the most prominent characteristics is the direct- to indirect-gap transition, which is also visible from Fermi surface reconstruction. 
		In high electron doping regime $n= 3.0 \times 10^{14}\,$ cm$^{-2}$    ($n= 4.0 \times 10^{14}\,$ cm$^{-2}$),  MoS$_2$ (WS$_2$) undergoes a direct-to-indirect gap transition under the one-sided gate doping. 
		Here, the $Q/Q^{\prime}$ valleys form the lower edge of the conduction band and are clearly larger in size than the $K/K^{\prime}$ pockets [Fig~\ref{fig:FMstructure} (b)]. 
		In the two-sided gate setup, the $K/K^{\prime}$ valleys form the conduction band edge so that they are correspondingly larger in size [Fig~\ref{fig:FMstructure} (c)].
		
%
%
		
	\end{appendix}	
	
	\newpage
	

%

\end{document}